\documentclass[a4paper,12pt]{article}
\pdfoutput=1 
\usepackage{jheppub} 

\usepackage[T1]{fontenc} 
\usepackage{physics,tensor,caption,subcaption}
\usepackage{float}

\def \cR {\mathcal{R}}

\def \Lstar {L_{\star}}

\title{Aspects of Geometric Inflation}

\author[a,b]{Jos\'e D. Edelstein,}
\author[a,b]{David V\'azquez Rodr\'\i guez,}
\author[a,b]{Alejandro Vilar L\'opez}
\affiliation[a]{Departamento de F\'\i sica de Part\'\i culas, Universidade de Santiago de Compostela, E-15782 Santiago de Compostela, Spain}
\affiliation[b]{Instituto Galego de F\'\i sica de Altas Enerx\'\i as (IGFAE), Universidade de Santiago de Compostela, E-15782 Santiago de Compostela, Spain}
\emailAdd{jose.edelstein@usc.es}
\emailAdd{davidvazquez.rodriguez@usc.es}
\emailAdd{alejandrovilar.lopez@usc.es}

\abstract{We revisit the recently proposed mechanism of Geometric Inflation. On general grounds, we show that obtaining the right amount of inflation demands an exceedingly large initial energy density. We introduce a scalar field and study the combined action of both mechanisms. Besides fixing the aforementioned issue, a cascading process occurs whose last step seems undistinguishable from ordinary large field inflation. Strikingly, the scalar field remains approximately constant while Geometric Inflation rules the dynamics. This ultimately leads to the possibility of reducing the initial value of the scalar field and its excursion. We discuss the main features of this hybrid scenario.}

\begin{document}
\maketitle

\section{Introduction}
\label{sec-Introduction}

It is a well-known fact that General Relativity (GR) in presence of standard fluids cannot explain an early stage of accelerated expansion in the universe. This led to the introduction of a scalar field, the inflaton, which in slow-roll conditions can trigger cosmological inflation. It was recently proposed, though, that higher curvature terms combined with some matter content have the potential to explain an early time inflationary epoch in the absence of an inflaton field \cite{Arciniega:2019oxa}. This is so even considering the energy-momentum tensor of such a conventional fluid as radiation, which would be unable to produce accelerated expansion in the context of GR. The mechanism was first proposed in the framework of cubic gravities \cite{Arciniega:2018fxj} and quickly generalized to include all orders \cite{Arciniega:2018tnn}. It was called Geometric Inflation and shown to provide a satisfactory inflationary evolution, at least at the background level.\footnote{The study of perturbations is quite involved in these theories, and definitely necessary to consider them as viable inflationary scenarios. Still, some partial results were derived in the cubic case \cite{Cisterna:2018tgx}, and generalized in \cite{EPVLunpublished}.}

A nice feature of these theories is that, order by order, each higher curvature density (i) possesses second-order linearized equations around any maximally symmetric background \cite{Bueno:2016xff}, (ii) admits generalizations of the Schwarzschild solution characterized by a single function, $g_{tt} g_{rr}=-1$, and whose thermodynamic properties can be accessed in a fully analytic fashion \cite{Hennigar:2016gkm, Bueno:2016lrh, Hennigar:2017ego, Bueno:2017sui}, and (iii) leads to a well-posed cosmological initial-value problem \cite{Arciniega:2018fxj, Arciniega:2018tnn}. It is a bottom-up construction relying on a conservative approach to a difficult and elusive problem. Interestingly enough, it was recently shown that a large class of higher curvature gravities ---those involving invariants constructed from contractions of the Riemann tensor and the metric and even some in which covariant derivatives are included--- are equivalent via field redefinitions to these theories \cite{Bueno:2019ltp}. More precisely, to theories complying the first two properties but not the cosmological one. It is conceivable that an analogous result including this latter property may be proven, although it is not clear at the moment.

Another nice feature comes from the fact that these theories provide a natural connection with late time evolution: radiation is the dominant component during inflation, where higher-curvature terms dominate and produce accelerated expansion, but then dilutes and the spacetime curvature decreases to the point where we can connect with a conventional GR epoch, where higher-curvature terms become irrelevant. The asymptotic structure of the phase space (of the cubic theory \cite{Arciniega:2018fxj}) reveals that an inflationary matter-dominated Big Bang is the global past attractor \cite{Quiros:2020uhr}. One can even obtain late time acceleration by turning on a cosmological constant, irrelevant during the inflationary epoch due to the strongly increasing behavior of $\rho_{\rm rad} \propto a^{-4}$ as the scale factor decreases. Further interesting results in this framework include \cite{Erices:2019mkd, Arciniega:2020pcy, Marciu:2020ysf}.

In spite of these attributes, there are good reasons to revisit the presence and the role of a scalar field in the early universe. On one hand, a recent study of odd-parity perturbations modes on a spatially homogeneous plane-symmetric Bianchi type I solution of the vacuum equations of motion in \cite{Arciniega:2018fxj}, showed the presence of (at least) one ghost which triggers a short-time-scale (compared to the Hubble time) classical instability \cite{Pookkillath:2020iqq}. This instability arises when a small anisotropy develops, and it appears to challenge the Geometric Inflation mechanism as a viable cosmological process. However, this analysis was performed in the cubic theory, around a de Sitter phase in the absence of radiation ---whose inclusion dramatically shifts from exponential to power-law accelerated expansion. It remains to be seen whether this strong instability remains once radiation is included and, most importantly, when the whole series of higher curvature terms is taken into account. The fast dilution of a barotropic fluid suggests that radiation may not be enough to cure this problem, unless exceedingly large values of its initial energy density are assumed. We will show that this is precisely the case and will commit ourselves to exploring the role of a scalar field in the context of Geometric Inflation.\footnote{A scalar field non-minimally coupled to Einsteinian Cubic Gravity \cite{Bueno:2016xff} ---which is not a cubic theory belonging to our class \cite{Arciniega:2018fxj}--- was recently considered \cite{Marciu:2020ski}.}

We will first show that the complete absence of a scalar field is phenomenologically and theoretically challenged. Besides the aforementioned arguments, it is not easy to foresee how to deal with reheating and how to reproduce the main features of cosmological perturbations in the absence of the inflaton. We show that obtaining the right amount of inflation demands an exceedingly large initial energy density for radiation. We will argue that the inclusion of a scalar field is demanded. However, it is easily seen that the accelerated expansion due to higher-curvature terms is boycotted by the presence of a scalar field. We will see, nevertheless, that the simultaneous inclusion of radiation and a scalar field has interesting consequences. For generic values of the couplings, the theory exhibits cosmological inflation with a remarkable feature: to a very good approximation, inflation happens in steps where the higher curvature terms rule initially, and there is a sort of cascade in which lower curvature terms follow higher curvature terms in governing the expansion, the last stage being undistinguishable from the familiar General Relativity setup with an inflaton slow-rolling field. That is, the familiar inflationary scenario becomes in this framework the very last step in a chain of successive stages.

We discuss the main features and some interesting prospects entailed by this scenario. First of all, in a generic situation the total number of e-folds splits into a Geometric Inflation phase followed by ordinary scalar field driven inflation. This points towards the simultaneous alleviation of both the radiation large initial energy density and the initial value and excursion of the scalar field, since the latter is not entirely responsible for the whole inflationary process. Besides, and strikingly, the scalar field remains pretty much constant while Geometric Inflation rules the cascading dynamics. The initial value of the scalar field and its excursion, though, remain transplanckian in this hybrid scenario. The fact that it always takes over in the last stage protects its role in the mechanism of reheating. Albeit the mechanism seems compelling, we will also present some criticisms and discuss some avenues worth exploring.

The article is organized as follows. In Section 2 we argue for the necessity of incorporating a scalar field in the realm of Geometric Inflation. Section 3 is devoted to presenting the conflict between a scalar field and higher curvature terms for the sake of producing cosmological inflation and showing how to deal with it. In Section 4 we study a framework in which there is both radiation and a scalar field. We analytically solve the problem in the regime where the $k$-th order term is dominant. We then numerically solve the full system and show to what extent the previous analytic solution is a good approximation. We see that inflation happens following a chain of different stages, which is neatly reflected in the behavior of the slow-roll parameter $\epsilon$. The familiar General Relativity setup becomes the last step at the end of the cascade. Section 5 is reserved for the discussion and conclusions.

\section{The need for a scalar field}
\label{sec-2}

Geometric Inflation has been built as a mechanism emerging from an {\it a priori} bottom-up construction where gravity is governed by
\begin{equation}
\mathcal{I}_{\rm GI} = \frac{1}{16\pi G} \int d^4 x \sqrt{-g} \left[ R + \sum_{n=3}^{\infty} c_n \Lstar^{2n-2} \cR_n \right] ~,
\label{action}
\end{equation}
where $\mathcal{R}_{(n)}$ is an $n$-th order density constructed from the Riemann tensor and its contractions satisfying a number of conditions.\footnote{It was recently shown that the cubic term entails causality violation through its contribution to the graviton $3$-point vertex in the Eikonal regime \cite{Camanho:2014apa} (see also \cite{Camanho:2015bbt}), which calls for the inclusion of an infinite tower of higher-spin particles with mass $\sim L_\star^{-1}$. This will not play a significant role in our approach for two reasons. Besides the obvious possibility of simply setting $c_3 = 0$, the tower of higher-spin particles would simply amount to another fluid sourcing the field equations.} First, it possesses second-order linearized equations of motion around any maximally symmetric background.\footnote{The restriction to maximally symmetric backgrounds is relevant. Lovelock's theorem guarantees that the only theory in four dimensions having second-order linearized equations of motion around \emph{any} background is General Relativity, possibly with a cosmological constant \cite{Lovelock:1971yv}. Imposing the condition only on maximally symmetric backgrounds implies that these theories share their spectrum with General Relativity: they only propagate a transverse and massless graviton in vacuum. More about this construction for general higher-order gravities can be found in \cite{Bueno:2016ypa}.} Second, it admits bona-fide non-hairy generalizations of the Schwarzschild black hole characterized by a single function, $g_{tt}g_{rr}=-1$, and whose thermodynamic properties can be accessed in a fully analytic fashion. Finally, and most importantly for the present work, it possesses a well-posed cosmological initial-value problem; namely, it admits cosmological Friedmann-Lema\^itre-Robertson-Walker (FLRW) solutions
\begin{equation}
ds^2 = -dt^2 + a(t)^2 \left(\frac{dr^2}{1-k r^2} + r^2 d\Omega^2 \right) ~,
\label{FLRW}
\end{equation}
where the associated generalized Friedmann equations for the scale factor $a(t)$ are second-order. In order to study cosmological inflation, it is more suitable to trade the time coordinate variable by the number of e-folds, $N$. In terms of the latter, the scale factor behaves as $a = \tilde{a}\,e^N$, where $\tilde{a}$ is the initial value; thereby\footnote{We implicitly assume $\dot{a} > 0$, which will certainly be the case for inflationary models.}
\begin{equation}
\frac{\mathrm{d}}{\mathrm{d}t} = H \frac{\mathrm{d}}{\mathrm{d}N} ~,
\label{dNefolds}
\end{equation}
where $H\equiv \dot{a}/a$ is the usual Hubble parameter. We use primes to denote derivatives with respect to $N$ and dots for the usual derivatives with respect to the cosmological time. The set of Friedmann equations reduce to a couple of second-order differential equations for the scale factor:
\begin{eqnarray}
3 F(H) &=& \frac{1}{M_{\rm Pl}^2} \rho ~, \label{FriedmannEq} \\ [0.4em]
- H^\prime\,\frac{dF(H)}{dH} &=& \frac{1}{M_{\rm Pl}^2} (\rho + P) ~,
\label{FriedmannEq2}
\end{eqnarray}
where $\rho$ and $P$ are the density and pressure obtained from the matter energy-momentum tensor, the reduced Planck mass is $M_{\rm Pl}^2 = (8\pi G)^{-1}$,
\begin{equation}
F(H) \equiv H^2 + \Lstar^{-2} \sum_{n=3}^{\infty} (-1)^n c_n \left( \Lstar H \right)^{2n} ~,
\label{F}
\end{equation}
and we have set $k = 0$. As usual, the conservation equation for the matter energy-momentum tensor, $\rho' + 3 (\rho + P) = 0$, is consistent with \eqref{FriedmannEq} and \eqref{FriedmannEq2}. The coefficients $c_n$ are expected to be computed from a UV complete theory of gravity.\footnote{For instance, T-duality is stringent enough to constrain all $\alpha^\prime$ corrections to the Friedmann equations (in the string frame) in terms of a single function which looks exactly like \eqref{F} \cite{Hohm:2019jgu}. String Theory is certainly invariant under $T$-duality ---among a broader set of theories \cite{Edelstein:2019wzg}---, thereby providing a rigorous scenario where the coefficients $c_n$ may be computed.} For the sake of exploration, we can also use a truncated function ---say, to the cubic or quartic order--- and set to zero all higher coefficients. We shall do that in this paper. It will allow us to show some features that would be less obvious in the general case, which we can address by straightforward extrapolation.

It was shown in \cite{Arciniega:2018tnn} that an accelerated expansion of the scale factor is a generic prediction of this theory if, for instance, the matter content is solely made of radiation. However, there are several reasons which could make the presence of extra fields contributing to the inflationary evolution beneficial. As an example, it is unclear how to deal with reheating in Geometric Inflation models, where the expansion is completely driven by an always diluting fluid. We will present here another argument which points towards the need of extra fields in the matter sector in order to make these models phenomenologically viable. Let us rewrite the first Friedmann equation \eqref{FriedmannEq} as
\begin{equation}
3 H^2 \left( 1 + G(H) \right) = \frac{1}{M_{\rm Pl}^2} \rho ~,
\label{GdeH}
\end{equation}
so that $G(H) = F(H)/H^2 - 1$ gives the higher-derivative contribution, isolating it from the GR part. In Geometric Inflation models, accelerated expansion appears when $G(H) \gg 1$, since within this higher-curvature domination regime normal fluids such as dust or radiation produce $\ddot{a} > 0$. Furthermore, the ability to gracefully connect with a late time evolution essentially identical to GR depends on the fact that $G(H) \to 0$ when $H \to 0$, and the decreasing behaviour of $H$, $H' \leq 0$.\footnote{With the second Friedmann equation \eqref{FriedmannEq2}, and considering that any fluid satisfying the null energy condition has $\rho + P \geq 0$, the decreasing evolution of $H$, $H' \leq 0$, is equivalent to the increasing nature of $F(H)$ as a function of $H$, $F'(H) \geq 0$.} Thus, we typically have two stages in these models: a first, higher-curvature dominated one, in which accelerated expansion happens and $G(H) \gg 1$, and a second, conventional one, where the evolution is essentially the same as in GR and where $G(H) \ll 1$. Let us define $H_{\rm end}$ as the value of the Hubble parameter for which $G(H_{\rm end}) = 1$, marking the scale of transition between the two stages.

The following will be a rough order of magnitude estimate, so we will not try to obtain exact expressions; it will nevertheless provide some insight to understand the problems of this set up. The energy density at the end of the accelerated expansion is $\rho_{\rm end} \sim M_{\rm Pl}^2 H_{\rm end}^2$. Assuming a fluid behavior, $\rho \sim a^{-3(1+ w)}$, with $w = 0$ for dust and $w = 1/3$ for radiation, if we allow for $N$ e-folds of inflation since a certain initial moment, the energy density at that initial time, $\rho_{\rm init}$, will be given by:
\begin{equation}
\rho_{\rm init} \sim \rho_{\rm end} e^{3(1+w)N} \sim M_{\rm Pl}^2 H_{\rm end}^2 e^{3(1+w)N} ~.
\label{rhoinit}
\end{equation}
It would be certainly tricky to allow for initial energy densities above the Planck scale. Demanding $\rho_{\rm init} < M_{\rm Pl}^4$, we obtain an upper bound on the value of $H$ at the end of the higher-curvature dominated stage:
\begin{equation}
H_{\rm end} \lesssim e^{-\frac{3}{2}(1+w)N} \, M_{\rm Pl} ~ .
\label{FluidGIBoundH}
\end{equation}
But $H_{\rm end}$ is setting the scale at which corrections to GR become noticeable in our gravitational theory. This is constrained by astrophysical tests. For the cubic theory, for instance, it was shown that Shapiro time delay experiments in the solar system lead to $\Lstar \lesssim 10^{8}$m \cite{Hennigar:2018hza}, assuming that $|c_3| \sim \mathcal{O}(1)$. This implies $H_{\rm end} \gtrsim 10^{-43} M_{\rm Pl}$. Similar results should be obtained for higher order theories as well since, in a sense, this is setting a lower energy bound for corrections to appear (it is reflecting the energy scale or, alternatively and maybe more clearly, the curvature length scale at which we are probing spacetime).\footnote{Indeed, analogous experiments in the context of the quartic theory were shown to lead to a constraint of the same order of magnitude \cite{Khodabakhshi:2020hny}.} This is a tiny fraction of the Planck energy, but the exponential dependence of \eqref{FluidGIBoundH} makes it quite relevant! In fact, for radiation and $N = 60$ we get $H_{\rm end} < 7.7 \times 10^{-53} M_{\rm Pl}$ in order to avoid super-Planckian energy densities at the beginning of inflation, which upsets the aforementioned observational constraints.\footnote{Notice, though, that dust ($\omega = 0$) can also drive inflation in these higher-curvature theories, and with $N = 60$ we obtain $H_{\rm end} < 8.2 \times 10^{-40} M_{\rm Pl}$, which is not in conflict with the observational constraint. However, the exponential dependence is still disturbing since increasing only to $N = 70$ we would get $H_{\rm end} < 2.5 \times 10^{-46} M_{\rm Pl}$.}

The previous discussion clearly shows one drawback of these Geometric Inflation models with a fluid driving all the evolution. The exponential expansion is so extreme that it needs tiny energy densities at the end of inflation in order for them to be below the Planck scale at the beginning. This in turn means that corrections to GR have to be relevant up until these tiny energy densities, otherwise accelerated expansion would stop in these models. But this conflicts with observational constraints on these corrections. Is it possible to get around this conclusion?

One obvious way is to consider inflation driven by something which does not dilute as fast as radiation or dust. In this sense, a scalar field is a good option: in a slow-roll regime it has $w \approx -1$ and therefore its energy density does not dilute at all. This would also make contact with the standard GR paradigm easier, since discussions concerning reheating can be more easily accommodated. In the following sections we will thus consider, as a simple model, different versions of the higher-curvature theories coupled to a free scalar field:
\begin{equation}
\mathcal{I}_{\rm scalar} = \int \mathrm{d}^4 x \sqrt{-g} \left[  - \frac{1}{2} \left( \nabla \phi \right)^2 - \frac{1}{2} m^2 \phi^2 \right] ~ .
\label{ScalarFieldAction}
\end{equation}
This matter sector contributes to the generalized Friedmann equations with the usual energy density and pressure:
\begin{equation}
\rho = \frac{1}{2} H^2 \phi'^2 + \frac{1}{2} m^2 \phi^2 ~, \qquad P = \frac{1}{2} H^2 \phi'^2 - \frac{1}{2} m^2 \phi^2 ~ ,
\label{ScalarFieldRhoAndP}
\end{equation}
and its equation of motion can be directly obtained from energy-momentum conservation:
\begin{equation}
\left( H \phi^\prime \right)^\prime + 3 H \phi^\prime + \frac{m^2}{H} \phi = 0  ~.
\label{ScalarEquation}
\end{equation}
%

\section{The inflaton and Geometric Inflation}
\label{sec-3}

Let us start by considering the simplest scenario, which is given by the cubic theory \cite{Arciniega:2018fxj} coupled to a scalar field with a quadratic potential \eqref{ScalarFieldAction},
\begin{equation}
S = \frac{M_{\rm Pl}^2}{2} \int d^4 x \sqrt{-g} \left[ R - \beta \Lstar^{4} \cR_3 \right] + \mathcal{I}_{\rm scalar} ~,
\label{cubicaction}
\end{equation}
where $c_3 = - \beta$ ($\beta > 0$) and
\begin{equation}
\begin{array}{cl}
\cR_3 & = - \displaystyle\frac{3}{4} R_{\alpha\ \beta}^{\ \gamma \ \delta} R_{\gamma\ \delta}^{\ \mu \ \nu} R_{\mu\ \nu}^{\ \alpha \ \beta} - \frac{1}{16} R_{\alpha\beta}^{\ \ \gamma\delta} R_{\gamma\delta}^{\ \ \mu\nu} R_{\mu\nu}^{\ \ \alpha\beta} + \frac{1}{2} R_{\alpha\beta\gamma\delta}R^{\alpha\beta\gamma}_{\ \ \ \mu}R^{\delta\mu} \\ [1.0em]
& \ - \displaystyle\frac{1}{6} R_{\alpha\beta\gamma\delta} R^{\alpha\beta\gamma\delta} R - \frac{1}{2} R_{\alpha}^{\beta} R_{\beta}^{\gamma} R_{\gamma}^{\alpha} - \frac{1}{4} R_{\alpha\beta\gamma\delta} R^{\alpha\gamma} R^{\beta\delta} + \frac{1}{4} R_{\alpha\beta} R^{\alpha\beta} R ~, \quad
\end{array}
\end{equation}
which is a unique linear combination, $\cR_3 = - \frac{1}{16} (\mathcal{P} - 8 \mathcal{C})$, of the Einsteinian Cubic Gravity action $\mathcal{P}$ \cite{Bueno:2016xff} and an independently characterized density $\mathcal{C}$ \cite{Hennigar:2017ego} leading to a well-posed cosmological initial value problem.

\begin{figure}[ht]
\centering\includegraphics[scale=1]{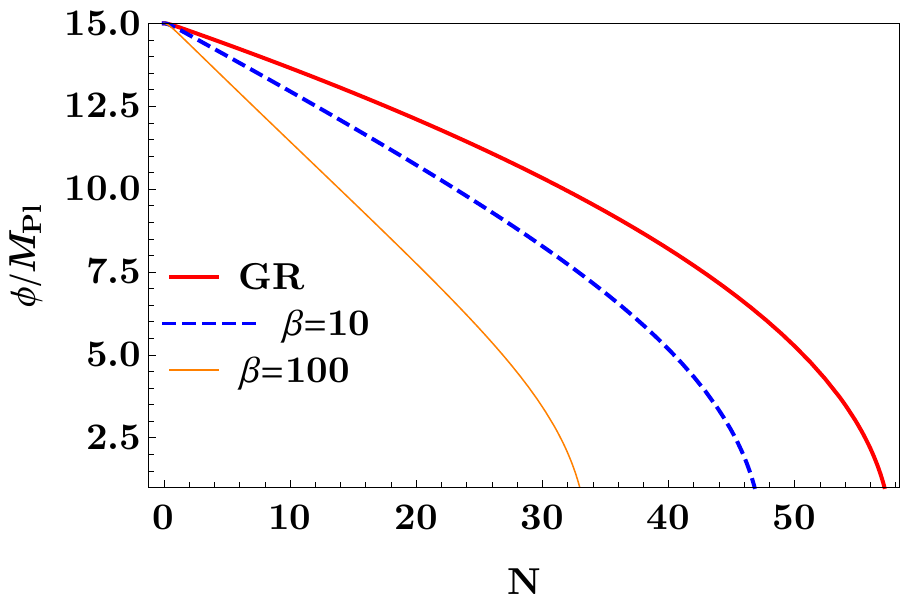}
\caption{Evolution of the scalar field for $\tilde\phi = 15 M_{\rm Pl}$, $m = 0.1 \Lstar^{-1}$ in Einstein gravity and cubic geometric inflation with $\beta=10$ and $\beta=100$. In General Relativity (red line) the scalar field stays almost constant during many e-folds, which is nothing but the slow-roll behavior, leading to $N_{\rm GR} \simeq 57$. In the cubic theory, instead, the scalar field drops quickly and $H_{\rm end}$ is reached very fast; then there is a brief GR inflationary period leading to $N_{\beta=10} \simeq 47$  and $N_{\beta=100} \simeq 33$ (blue dotted line and orange line respectively). The end of inflation, as usual, is taken as the value of $N$ at which the slow-roll parameter $\epsilon = - \dot{H}/H^2 = - H^\prime/H = 1$.}
\label{CubicLessEfficient}
\end{figure}

Consider the situation in which a scalar field with initial value $\tilde{\phi}$ and the usual quadratic potential is left to evolve on its own. The equations of motion are \eqref{ScalarEquation} and
\begin{equation} 
3 H^2 \left( 1 + \beta H^4 \right) = \frac{1}{2} H^2 {\phi^\prime}^2 + \frac{1}{2} m^2 \phi^2 ~,
\label{FriedmannEquationCubic_Scalar}
\end{equation}
where all units have been absorbed: $H$ and $m$ are given in units of $\Lstar^{-1}$, while $\phi$ is written in Planck units. It is then easy to check that for fixed values of $\tilde\phi$ and $m$ the cubic theory is less efficient as an inflationary theory; see Figure \ref{CubicLessEfficient}.
If we reduce the value of $m$ while keeping $\tilde\phi$ fixed, the number of e-folds in the cubic theory grows and approaches the result of General Relativity; this is simply telling us that $H_{\rm end}$ occurs earlier (since the evolution starts from a lower $H_{\rm init}$), and most of the dynamics is governed by the Einstein-Hilbert term.

The situation changes if we bring in radiation into the picture. The Friedmann equation acquires a new contribution
\begin{equation} 
3 H^2 \left( 1 + \beta H^4 \right) = \tilde{\rho}\,e^{-4N} + \frac{1}{2} H^2 {\phi^\prime}^2 + \frac{1}{2} m^2 \phi^2 ~,
\label{FriedmannEqRadiation}
\end{equation}
where $\tilde{\rho} = \xi^2 M_{\rm Pl}^{-4} \rho_{\rm init}$ ---we introduce for convenience $\xi = \Lstar M_{\rm Pl}$, the ratio between $\Lstar$ and the Planck length---, while the continuity equation \eqref{ScalarEquation} remains the same. Radiation energy density decays pretty fast irrespective of what is the mechanism driving inflation. This is the reason behind the decelerated expansion it entails in General Relativity. Now, the higher curvature dynamics is relevant at earlier times, on generic grounds, where also the ratio between radiation and scalar field energy densities is large. Besides, we know that inflation driven by the scalar field happens in a slow-roll regime, and the decay of its energy density should be slower than that of radiation. A possibility arises: is there a range of parameters for which inflation is a combined effect of Geometric Inflation driven by radiation followed by the familiar slow-roll setup of a scalar field in GR? For this to be the case, the scalar field must remain approximately constant while inflation is driven by radiation, and $H_{\rm end}$ must take place not far from the moment at which the energy densities are commensurate. We will see that the answer is yes, and the range of parameters is broad enough not to be considered the effect of fine tuning.

Notice that this scenario could have been spoiled in many ways. For instance, if the radiation expansion gives kinetic energy to the scalar it could be impossible to enter the second stage dominated by the scalar field in a slow-roll regime, which would mean that it does not trigger inflation. Or it could be that the two stages are not truly decoupled and in the transition from radiation domination to scalar field domination weird effects make the accelerated expansion impossible. We will show that this is not the case, but in order to do that it will prove useful to analyze both regimes of the expansion independently, under the decoupling assumption. The extremely different behavior of the radiation and the scalar field energy densities leads us to conjecture that the expansion in each of these stages is dominated by a single component. We will see that this is a good approximation.

\subsection{Radiation domination}
\label{subsec-3-1}

Consider then a period of expansion where radiation dominates the right-hand side of \eqref{FriedmannEqRadiation} and the cubic term the left-hand side,
\begin{equation} 
H(N) \approx H_{\rm init}\,e^{-2N/3} ~,
\label{RadiationDominates}
\end{equation}
where $H_{\rm init} = \sqrt[6]{\tilde\rho/(3\beta)}$. This regime stops either when the GR term equals the cubic one ---{\it i.e.}, when there is a value of $N = N_{\rm GR}$ for which $H(N_{\rm GR}) = H_{\rm end}$---, or when the scalar field energy density becomes the dominant contribution ---{\it i.e.}, when $N = N_{\rm s}$ such that, in a slow-roll regime, $\frac12 m^2 \phi(N_{\rm s})^2 = \tilde{\rho}\,e^{-4 N_{\rm s}}$ (whose expression \eqref{Nscalar} is derived below)---, whatever comes first. The expression for $N_{\rm GR}$ is given by
\begin{equation} 
N_{\rm GR} = \frac{1}{8} \log \left( \frac{\beta \tilde{\rho}^2}{9} \right) ~.
\label{eFoldsCubicDominationRadiation}
\end{equation}
since $\beta H(N_{\rm GR})^4 = 1$. On the other hand, in order to compute when the scalar field overwhelms radiation, we need to know its evolution under radiation domination, therefore, when the Hubble parameter behaves as in \eqref{RadiationDominates}. The scalar field equation \eqref{ScalarEquation} can be written in this case as
\begin{equation}
\phi^{\prime\prime}(N) + \frac{7}{3} \phi^\prime(N) + \left( \frac{m}{H_{\rm init}} \right)^2 e^{4N/3}\,\phi(N) = 0 ~.
\label{ScalarFieldEquationCubicRadiation}
\end{equation}
Under the following redefinition
\begin{equation}
\phi(N) \equiv \left( \frac{m}{H_{\rm init}} e^{2N/3} \right)^{-7/4} f\left( \frac{3m}{2 H_{\rm init}} e^{2N/3} \right) ~,
\label{phiandf}
\end{equation}
it can be seen that the function $f$ satisfies the following Bessel equation: 
\begin{equation}
x^2 \frac{\mathrm{d}^2\!f(x)}{\mathrm{d} x^2} + x \frac{\mathrm{d} f(x)}{\mathrm{d} x} + \left[ x^2 - \left(\frac{7}{4} \right)^2 \right] f(x) = 0 ~.
\label{BesselEquationCubic}
\end{equation}
The most general solution for the scalar field employing Bessel and Neumann functions is thus
\begin{equation}
\phi(N) = e^{-7N/6} \left[ C_1\,J_{7/4} \left( \frac{3m}{2 H_{\rm init}} e^{2N/3} \right) + C_2\,N_{7/4} \left( \frac{3m}{2 H_{\rm init}} e^{2N/3} \right) \right] ~,
\label{ScalarFieldEvolutionCubicRadiation}
\end{equation}
where we absorbed some pre-factors into the integration constants. There are a couple of interesting facts hidden in the previous equation. Bessel functions for large argument, $x \gg 1$, behave asymptotically as oscillating trigonometric functions with a suppressing factor:
\begin{equation}
J_{7/4}(x) \sim \sqrt{\frac{2}{\pi x}} \cos \left( x - \frac{9 \pi}{8} \right) ~ , \quad N_{7/4}(x) \sim \sqrt{\frac{2}{\pi x}} \sin \left( x - \frac{9 \pi}{8} \right) ~.
\label{BesselAsymptotics}
\end{equation}
This means that we have an exponential suppression of the scalar field
\begin{equation}
\phi(N) \sim e^{-3N/2} ~, \qquad {\rm for} \quad N \gg N_\star = \frac{3}{2} \log \left( \frac{2 H_{\rm init}}{3m} \right) ~.
\label{ScalarFieldSuppression}
\end{equation}
Note that in this regime the potential energy density scales as $\phi^2 \sim e^{-3N}$, which is the expected result for dust, and exponentially slower than the radiation energy density, as anticipated. If we want to enter a second phase of inflation driven by a slow-rolling scalar field in GR, we need to avoid the previous exponential suppression. The number of e-folds spent in the radiation domination phase must necessarily be lower or equal than $N_\star$.

The physically interesting scenario involves a phase of Geometric Inflation with a reasonable share of the early universe expansion, say, at least $10$ e-folds; thereby we consider $3m/(2 H_{\rm init}) \ll 1$. Then, it can be shown that setting initial conditions $\phi(0) = \tilde{\phi}$ and $\phi^\prime(0) = 0$ is equivalent to cancelling $C_2$ and fixing $C_1$ so that\footnote{This was verified rigorously through a series expansion in $m/H_{\rm init}$ and using the asymptotics of Bessel functions. Nevertheless, it can be understood recalling the singular behavior of Neumann functions at the origin, which forces us to cancel the contribution of $N_{7/4}$ when $m/H_{\rm init} \ll 1$ in order to ensure regularity at $N=0$.}
\begin{equation} 
\phi(N) \approx \Gamma \left( 11/4 \right) \left( \frac{4 H_{\rm init}}{3 m} \right)^{7/4} e^{-7N/6}\, J_{7/4} \left( \frac{3m}{2 H_{\rm init}} e^{2N/3} \right)\,\tilde{\phi} ~.
\label{ApproximationScalarCubicRadiation}
\end{equation}
This is the last bit of information we needed to determine when the scalar field energy density becomes the dominant component. Assuming that the potential energy is much larger than the kinetic energy ---in order to be in a slow-roll inflationary epoch---, the scalar field and radiation energy densities become commensurate for $N_{\rm s}$, such that $\frac12 m^2 \phi(N_{\rm s})^2 = \tilde{\rho}\,e^{-4 N_{\rm s}}$; thereby
\begin{equation}
N_{\rm s} = \frac{3}{2} \log \left( \frac{2 H_{\rm init}}{3 m} x_{\rm s} \right) ~,
\label{Nscalar}
\end{equation}
where $x_{\rm s}$ is the value fulfilling 
\begin{equation} 
x_{\rm s}^{5/2} J_{7/4}(x_{\rm s})^2 = \frac{3^7}{2^{9/2} \, \Gamma(11/4)^2} \frac{m^4 \beta}{16 \, \tilde{\phi}^2} ~.
\label{xEquationCubicRadiation}
\end{equation}
If $x_{\rm s} \ll 1$, this can be further simplified through a series expansion of the Bessel function; we obtain
\begin{equation}
N_{\rm s} \approx \frac{1}{4} \log \left( \frac{2 \tilde{\rho}}{m^2 \tilde{\phi}^2} \right) ~.
\label{ApproximationNSCubicRadiation}
\end{equation}
Notice that if $N \ll N_\star$,
\begin{equation}
\Delta\phi \equiv \tilde{\phi} - \phi(N) \sim \tilde{\phi} \, e^{4(N - N_\star)/3} ~,
\label{RoughlyConstant}
\end{equation}
and we conclude that the scalar field stays roughly constant! Essentially, radiation drives a very fast expansion during which the scalar field remains constant, guaranteeing that if initial conditions are chosen to allow slow-roll evolution, the radiation domination phase will not spoil them. This seems to us remarkable and unexpected.

\subsection{Inflaton domination}
\label{subsec-3-2}

\begin{figure}[ht]
\centering \includegraphics[scale=1.00]{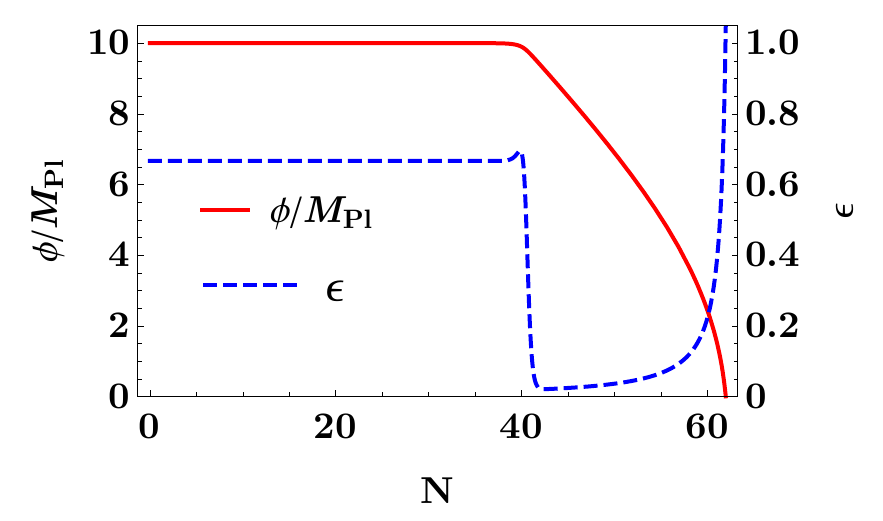} 
\caption{The scalar field stays pretty much constant during the $40$ e-folds of Geometric Inflation and then slowly rolls down the potential leading to a total of about $60$ e-folds. The slow-roll $\epsilon$-parameter starts at $2/3$, the value for radiation in the cubic theory \cite{Arciniega:2018fxj}, and when the slowly rolling scalar field starts to dominate it drastically drops, displaying a quasi-de Sitter behavior not different from the GR case, albeit with a lower value of the scalar field.}
\label{FigureEpsilonFTCubic}
\end{figure}

We will consider the most efficient situation in which the first epoch of inflation finishes both when the scalar field starts to dominate and General Relativity dynamics takes over the cubic term; that is, we impose $N_{\rm GR} = N_{\rm s}$. From \eqref{eFoldsCubicDominationRadiation} and \eqref{Nscalar},
\begin{equation} 
\frac{1}{6} \sqrt{\beta}\,m^2 \tilde{\phi}^2 = 1 ~;
\label{FineTunedMass}
\end{equation}
this fixes the mass of the scalar once its initial value is known. Let us remark that this is not a necessary condition but rather an optimal one. The second part of the inflationary evolution is well-known. The scalar field slowly rolls down the potential producing an accelerated expansion governed by the Einstein-Hilbert term, provided that the potential energy dominates over the kinetic energy. If $\phi_{\rm i}$ is the initial value of the scalar field, the number of e-folds of inflation produced reads
\begin{equation} 
N(\phi_{\rm i}) = \frac{\phi_{\rm i}^2}{4} - \frac{1}{2} ~.
\label{eFoldsScalarGR}
\end{equation}
Since $\beta$ is expected to be $\mathcal{O}(1)$, let us set for definiteness $\beta = 1$. For a setup in which Geometric Inflation is responsible for $40$ e-folds, $N_{\rm GR} = 40$, we see that we would need $\phi_{\rm i} \simeq 9 M_{\rm Pl}$ in order to complete the $60$ e-folds for the whole process.
The initial value of the scalar field $\tilde\phi$ is a bit larger than this; {\it e.g.}, in Figure \ref{FigureEpsilonFTCubic} we ran a numerical evolution with $\tilde\phi = 10 M_{\rm Pl}$.
Then \eqref{FineTunedMass} fixes $m = 0.24 \Lstar^{-1}$. 

Notice that this resolves the issue with the initial radiation energy density discussed in the previous Section. In this particular example, for instance, we get $\tilde\rho = \xi^2 M_{\rm Pl}^{-4} \rho_{\rm init} \simeq 9.2 \times 10^{69}$, and the current constraint $\xi < 10^{43}$ \cite{Hennigar:2018hza} leaves plenty of room for subplanckian values of $\rho_{\rm init}$. The shorter the period of Geometric Inflation the better in terms of this issue.

Let us verify that the different approximations are justified. We do not get an exponential suppression of the scalar field (and in fact it stays roughly constant during radiation domination) because of the values of the following two quantities:
\begin{equation}
\frac{3 m}{2 H_{\rm init}} = 8.59 \times 10^{-13} ~, \qquad x_{\rm s} = \frac{3 m}{2 H_{\rm init}} e^{2 N_{\rm GR}/3} = 0.37 ~.
\end{equation}
This is also the reason guaranteeing the legitimacy of cancelling the contribution to the inflaton coming from the Neumann function. Also, it is the reason behind taking the approximation \eqref{ApproximationNSCubicRadiation} instead of the full equation involving the Bessel function. Although $x_{\rm s} \ll 1$ is not true, the first term of the series is still good enough. Finally, $\tilde{\rho} \gg m^2 \tilde{\phi}^2/2$ constitutes a trivial check that indeed we start in a radiation dominated phase.

\section{Inflation at the end of the cascade}
\label{sec-4}

We would like to consider the inclusion of higher curvature terms. We will see that the general set-up remains the same: we aim for a radiation dominated initial phase of Geometric Inflation where higher curvature terms are relevant, followed by a last stage where the scalar field starts to rule the dynamics and gravity is essentially governed by the Einstein-Hilbert term. Under quite general conditions we find a cascading dynamics with a sequence of stages where decreasing higher curvature terms dictate the universe expansion until the very last step where ordinary inflation takes place. We will explore how this scenario alleviates the issues related to the large values of radiation energy density and the transplanckian initial value of the scalar field.

Consider a system governed by $\mathcal{I}_{\rm GI}$ \eqref{action}, coupled to a free massive scalar field and radiation, whose Euler-Lagrange equations can be summarized in the Friedmann equation
\begin{equation}
3 F(H) = 3 H^2 \left( 1 + \sum_{n=3}^{\infty} (-1)^n c_n H^{2(n-1)} \right) = \tilde{\rho}\, e^{-4N} + \frac{1}{2} H^2 {\phi^\prime}^2 + \frac{1}{2} m^2 \phi^2 ~,
\label{GeneralizedFriedmannEquation}
\end{equation}
and the conservation of the scalar field energy-momentum tensor \eqref{ScalarEquation}. The cubic case presented in the previous Section would be given by $c_3 = - \beta$, and all the remaining coefficients vanishing. We want to understand the dynamics driven by radiation and the $p$-th order term. We assume that such term neatly dominates the behavior of the system until the Hubble parameter decreases enough such that the $(p - 1)$-th term becomes the relevant one, and so on and so forth.\footnote{A comment is in order. This assumption is unnatural if we assume that \eqref{action} is an Effective Field Theory, in which case we would expect all terms being relevant at once for processes involving energies $E \sim \Lstar^{-1}$. However, since ours is a bottom-up construction, we can also explore phenomenologically the case in which just a few $c_n$ are turned on, and compare with the experimental data. On more pragmatical grounds, it is easier to deal with a truncated series and infer the consequences of our analysis once all terms in \eqref{action} are included.} We will end up by numerically solving a quartic example in which we shall see the extent to which this assumption is well supported by the actual data. So let us analyze what happens in a radiation dominated phase where gravity is essentially controlled by the $p$-th term in the series. The Hubble parameter would evolve as
\begin{equation}
H(N) = H^{(p)}_{\rm init}\, e^{-2N/p} ~, \qquad {\rm where} \quad H^{(p)}_{\rm init} = \left(  \frac{\tilde\rho_p}{3 c_p} \right)^{\frac{1}{2p}} ~,
\label{HubbleParameterPGravity}
\end{equation}
$\tilde\rho_p$ is the initial value of the radiation energy density in the step where the $p$-th order term dominates. Similarly, $H^{(p)}_{\rm init}$ is the initial value of the Hubble parameter. To be more concrete, in this paper we will consider a truncated action where there is a maximum value, $p = n_{\rm max}$. In that case, $\tilde\rho_{n_{\rm max}} = \tilde\rho$, while $\tilde\rho_{n_{\rm max}-1} = \tilde\rho\, e^{-4 N_{n_{\rm max}}}$, where the dilution experienced after an expansion of $N_{n_{\rm max}}$ e-folds corresponding to that step in the cascade is manifest. The slow-roll $\epsilon$-parameter acquires a definite value at each step:
\begin{equation}
\epsilon(N) = - \frac{H^\prime(N)}{H(N)} = \frac{2}{p} ~,
\end{equation}
and we have an accelerated expansion for $p \geq 3$. The $p$-th order term stops being the dominant one and we enter into the realm of the $(p-1)$-th order domination after $N_p$ e-folds of expansion,\footnote{Notice that we are assuming $c_{p-1} \neq 0$, otherwise we would have to consider the change from a $p$-th order term to a $(p-k)$-th order term for $k \neq 1$.}
\begin{equation}
N_p = \frac{1}{4} \log \left( \frac{\tilde\rho_p}{3}\,\frac{c_p^{p-1}}{c_{p-1}^p} \right) ~,
\label{PGravityToPm1}
\end{equation}
where the following condition holds: $c_p H(N_p)^{2(p-1)} = c_{p-1} H(N_p)^{2(p-2)}$. We can now consider how the scalar field evolves in this radiation dominated background, just as we did in the cubic case. The equation \eqref{ScalarEquation} can be written as
\begin{equation}
\phi^{\prime\prime}(N) + \frac{3p-2}{p} \phi^\prime(N) + \left( \frac{m}{H^{(p)}_{\rm init}} \right)^2 e^{4N/p}\,\phi(N) = 0 ~,
\label{ScalarFieldEquationRadiationBackgroundPGravity}
\end{equation}
and it can be solved in terms of Bessel functions:
\begin{equation*}
\phi(N) = e^{-\frac{(3p -2)N}{2p}} \left[ C_1 \, J_{(3p-2)/4} \left( \frac{p}{2} \frac{m}{H^{(p)}_{\rm init}} \, e^{2N/p} \right) + C_2 \, N_{(3p-2)/4} \left( \frac{p}{2} \frac{m}{H^{(p)}_{\rm init}} \, e^{2N/p} \right) \right] ~ .
\label{ScalarBesselFunctionsPGravity}
\end{equation*}
All the discussions presented in the cubic case are directly translated into this general case: we need $m/H^{(p)}_{\rm init} \ll 1$ in order to have a number of e-folds of expansion without the suppression due to the scalar field, etcetera. Notice that it is enough to fulfill this inequality for the last stage, since $H^{(p\neq 3)}_{\rm init} > H^{(3)}_{\rm init}$. Assuming also the scalar field to start without velocity, and this condition to be kept in the successive steps of the cascade, while radiation dominates the expansion,
\begin{equation}
\phi(N) \approx e^{-\frac{(3p -2)N}{2p}} \left( \frac{4}{p} \frac{H^{(p)}_{\rm init}}{m} \right)^{(3p-2)/4}\!\!\!\!\!\Gamma \left( (3p+2)/4 \right)\, J_{(3p-2)/4} \left( \frac{p}{2} \frac{m}{H^{(p)}_{\rm init}} \, e^{2N/p} \right) \tilde\phi_p ~,
\label{ApproximationScalarBesselFunctionsPGravity}
\end{equation}
where $\tilde\phi_p$ is the initial value in the regime where the $p$-th order term governs the dynamics. $N$ is here the number of e-folds of expansion since we entered the $p$-th order phase as well. To leading order in the Bessel function argument, the previous equation is a constant equal to $\tilde\phi_p$ ---and this is the case all along the cascade---. Remarkably, the scalar field remains pretty much constant, as a spectator of the different phases of Geometric Inflation.

For completeness, we can equate the scalar field potential energy density to the radiation energy density in order to find out when the scalar field would take control over the evolution in the $p$-th order phase.\footnote{In practice, we do not want to do this in any concrete step but the latest, $p=3$, since, as discussed above, General Relativity is the most efficient theory inflating with a scalar field.} This happens for a number of e-folds $N^{(p)}_{\rm s}$ determined by solving
\begin{equation}
x_{\rm s}^{(p+2)/2} J_{(3p-2)/4} (x_{\rm s})^2 = \frac{3 p^{2p}}{2^{(7p-4)/2}} \frac{c_p}{\Gamma \left( (3p+2)/4 \right)^2} \frac{m^{2(p-1)}}{\tilde\phi_p^2} ~,
\label{xSEquationPGravity}
\end{equation}
and then replacing
\begin{equation}
N^{(p)}_{\rm s} = \frac{p}{2} \log \left( \frac{2}{p} \frac{H^{(p)}_{\rm init}}{m} x_{\rm s} \right) ~.
\end{equation}
Let us remind an important feature of this construction: we can keep a scalar field essentially constant during the sequence of radiation dominated phases provided that $m/H^{(p)}_{\rm init} \ll 1$ for all the steps in the cascading evolution.

\begin{figure}[ht]
\centering\includegraphics[scale=0.7]{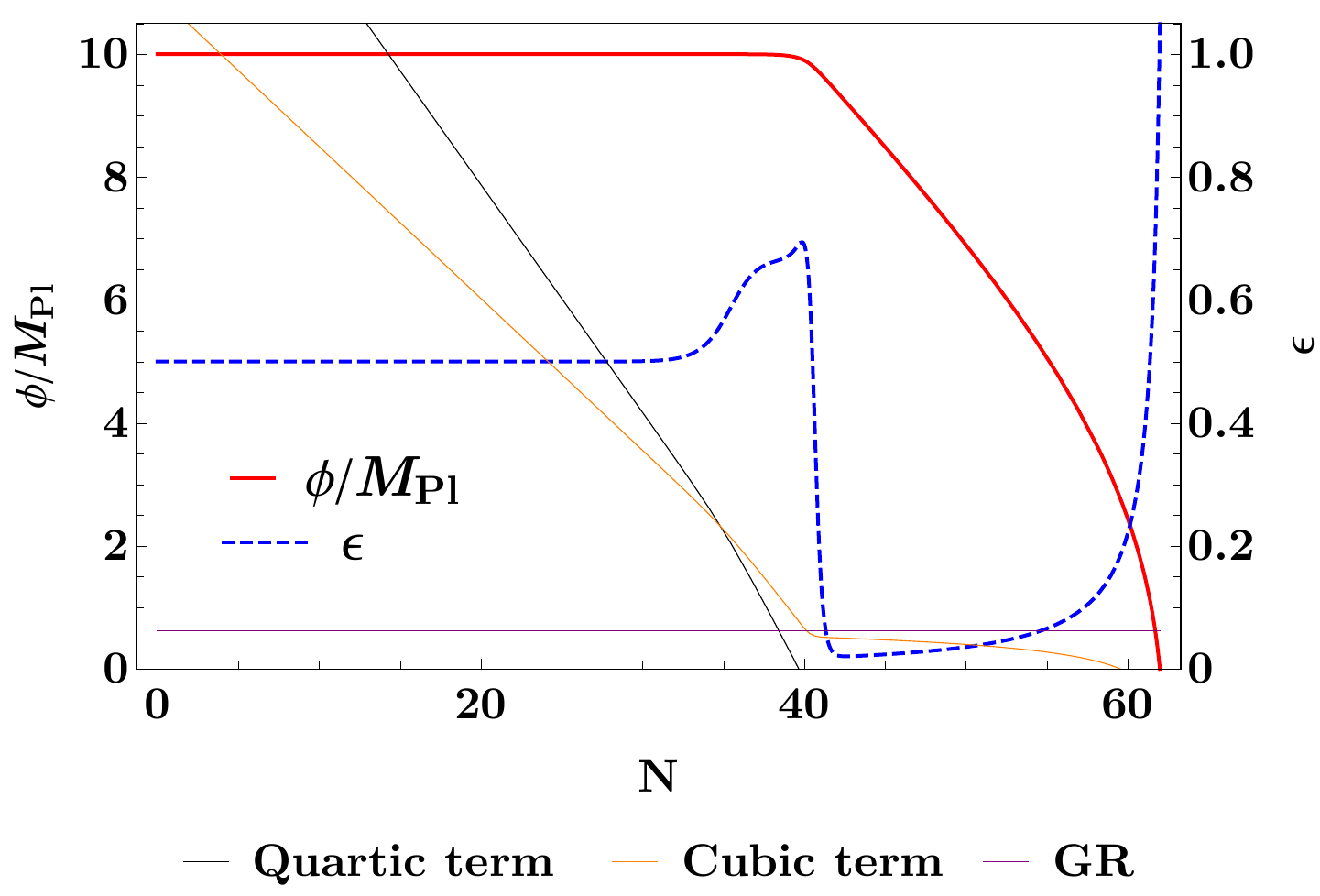}
\caption{Comparison between the different contributions inside the parenthesis of the left hand side of \eqref{ForFigure3} (thin lines) in logarithmic scale, displaying the subsequent dominance from higher to lower curvature terms. The numerical solution shows that separation in phases is a good approximation. The slow-roll $\epsilon$-parameter (dotted blue line) starts at $1/2$, then jumps to $2/3$, corresponding to the expected behavior of radiation driven expansion in the quartic and cubic theories, and when the slowly rolling scalar field dominates it drops, displaying a quasi-de Sitter behavior not different from the GR case, albeit with a lower value of the scalar field (thick red line).}
\label{FigureEpsilonFTCuartic}
\end{figure}

Let us end by considering an example in order to test our assumptions against the actual numerical solution. We add a quartic term to the cubic theory presented in the previous Section. We shall call $c_4 = \lambda$, and for the numerical computations we will stick to the previously chosen value $\beta = 1$. The Friedmann equation becomes
\begin{equation}
3 H^2 \left( 1 + \beta H^4 + \lambda H^6 \right) = \tilde{\rho}\, e^{-4N} + \frac{1}{2} H^2 {\phi^\prime}^2 + \frac{1}{2} m^2 \phi^2 ~.
\label{ForFigure3}
\end{equation}
If our assumption about the scalar field staying constant during the radiation dominated evolution is correct, we can choose $\tilde{\phi} = 10 M_{\rm Pl}$ and $m = 0.24 \Lstar^{-1}$, as in the cubic case, so that we would have $20$ e-folds of scalar inflation in General Relativity and the transition to inflaton domination will occur when the cubic term ceases to overwhelm the Einstein-Hilbert contribution. The initial value of the radiation energy density only depends on the number of e-folds along the Geometric Inflation cascade. In this example, since we are keeping the values used in the cubic example, if the total number of e-folds is kept fixed, so will be $\tilde\rho$. The number of e-folds correspondent to the quartic epoch can be obtained from \eqref{PGravityToPm1},
\begin{equation}
N_4 = \frac{1}{4} \log \left( \frac{\tilde\rho}{3}\,\frac{\lambda^{3}}{\beta^4} \right) ~.
\end{equation}
For instance, if $\lambda = 0.001$, we would obtain $N_4 \simeq 35$, a value that leads to a scenario in which the $60$ e-folds are roughly divided into three pieces. This is plotted in Figure \ref{FigureEpsilonFTCuartic}, where the three regions for which the different terms dominate become apparent.
Notice also that the regime led by the cubic term, where the expansion develops about $5$ e-folds, is not as neat as the quartic stage. The length ---in e-folds--- of each step in the cascade depends upon the relative values of the coefficients $c_n$. Higher curvature terms tend to rule the expansion during the majority of the $N_{\rm GR}$ e-folds. Only if $c_n \ll c_{n-1}$, the cascade displays the different stages as an open bellow or bandoneon, consisting of a series of power-law stages smoothly connected with a last stage where slow-roll inflaton in GR takes over. Otherwise, for commensurate values of the $c_n$ parameters, $c_n \sim \mathcal{O}(1)$, the higher curvature contribution takes over all the lower curvature ones, which drastically shrink to zero size.

\section{Discussion and conclusions}
\label{sec-5}

Let us first summarize our findings. We have shown, on general grounds, that the Geometric Inflation mechanism purely driven by radiation demands a transplanckian initial energy density. We then included a scalar field with a quadratic potential as a case study. We first saw that inflation driven by a slowly rolling scalar field in General Relativity is more efficient than in the higher curvature counterparts. When we included radiation into the picture, however, for a wide range of parameters we saw that there is a cascading scenario in which Geometric Inflation rules first, the dominance being transferred from higher to lower curvature terms, the last stage being the familiar regime of General Relativity in the slow-roll approximation.

This is manifest provided that the different couplings of the higher-order terms satisfy some hierarchical relations allowing each of them to dominate separately. Albeit this would be unnatural from an Effective Field Theory perspective, in which case we would expect all terms being relevant at once for processes involving energies $E \sim \Lstar^{-1}$, ours is a bottom-up construction to be understood as a phenomenological exploration in which just a few $c_n$ may be turned on. On more pragmatical grounds, it is easier to deal with a truncated series and infer the consequences of our analysis once all terms in \eqref{action} are included. Strikingly enough, the scalar field remains almost frozen while the universe undergoes Geometric Inflation. Besides, given the standard aspect of the last inflationary stage, we do not expect modifications regarding the reheating paradigm \cite{Weinberg:2008zzc}.

Within the current framework we saw that it is possible to reduce the energy density of radiation and marginally solve its exceedingly huge value in Geometric Inflation. By marginally we mean that our resolution relies on the current constraints on $\xi = \Lstar M_{\rm Pl}$, which one may foresee may be drastically improved in the future. On the other hand, we could lower the initial value of the scalar field but it is still within the superplanckian range. The very fact that the cascade ultimately enters in the GR regime bounds us to obey $\tilde\phi \geq \sqrt{2} M_{\rm Pl}$ if $N(\phi_{\rm i}) > 0$ \eqref{eFoldsScalarGR}, since $\tilde\phi \gtrsim \phi_{\rm i}$. The quadratic potential still belongs to the large field inflationary models. The question arises whether it is possible to have a situation in which cosmological inflation fully happens in the high curvature regime in the absence of radiation. The answer is in the affirmative and it is possible to convert a scalar field with a quadratic potential, a benchmark of the large field models, into a small field inflationary model \cite{Edelstein:2020lgv}.

The study of cosmological perturbations in this framework is a difficult task \cite{EPVLunpublished}. However, the separation between Geometric Inflation and the GR regime makes it a little more feasible. An interesting open problem is to prove that the current hybrid scenario is free from the strong instabilities presented in \cite{Pookkillath:2020iqq}. We expect to report on some of these issues in the near future.

\section*{Acknowledgements}

We have benefitted from discussions with Jos\'e Juan Blanco Pillado, Pablo Cano, Antonio De Felice, Robert Mann, Masroor Pookkillath, Alberto Rivadulla, Alexei Starobinsky, Gianmassimo Tasinato and Ivonne Zavala. We are specially grateful to Sonia Paban for her comments and collaboration on the subject.
This work is supported by MINECO FPA2017-84436-P, Xunta de Galicia ED431C 2017/07, FEDER, and the Mar\'\i a de Maeztu Unit of Excellence MDM-2016-0692. DVR is supported by Xunta de Galicia under the grant ED481A-2019/115. AVL is supported by the Spanish MECD fellowship FPU16/06675.
DVR and AVL are respectively pleased to thank the University of Waterloo and the University of Texas at Austin, where part of this work was done, for their warm hospitality.



\begin{thebibliography}{99}

\bibitem{Arciniega:2019oxa}
G.~Arciniega, P.~Bueno, P.~A.~Cano, J.~D.~Edelstein, R.~A.~Hennigar and L.~G.~Jaime, ``Cosmic inflation without inflaton,''
Int. J. Mod. Phys. D \textbf{28}, 1944008 (2019).

\bibitem{Arciniega:2018fxj}
G.~Arciniega, J.~D.~Edelstein and L.~G.~Jaime, ``Towards geometric inflation: the cubic case,'' Phys. Lett. B \textbf{802}, 135272 (2020) [arXiv:1810.08166 [gr-qc]].

\bibitem{Arciniega:2018tnn}
G.~Arciniega, P.~Bueno, P.~A.~Cano, J.~D.~Edelstein, R.~A.~Hennigar and L.~G.~Jaime, ``Geometric Inflation,'' Phys. Lett. B \textbf{802}, 135242 (2020) [arXiv:1812.11187 [hep-th]].

\bibitem{Cisterna:2018tgx}
A.~Cisterna, N.~Grandi and J.~Oliva, ``On four-dimensional Einsteinian gravity, quasitopological gravity, cosmology and black holes,'' Phys. Lett. B \textbf{805}, 135435 (2020) [arXiv:1811.06523 [hep-th]].

\bibitem{EPVLunpublished}
J.~D.~Edelstein, S.~Paban and A.~Vilar L\'opez, in progress.

\bibitem{Bueno:2016xff}
P.~Bueno and P.~A.~Cano, ``Einsteinian cubic gravity,'' Phys. Rev. D \textbf{94}, 104005 (2016) [arXiv:1607.06463 [hep-th]].

\bibitem{Hennigar:2016gkm}
R.~A.~Hennigar and R.~B.~Mann, ``Black holes in Einsteinian cubic gravity,'' Phys. Rev. D \textbf{95}, 064055 (2017) [arXiv:1610.06675 [hep-th]].

\bibitem{Bueno:2016lrh}
P.~Bueno and P.~A.~Cano, ``Four-dimensional black holes in Einsteinian cubic gravity,'' Phys. Rev. D \textbf{94}, 124051 (2016) [arXiv:1610.08019 [hep-th]].

\bibitem{Hennigar:2017ego}
R.~A.~Hennigar, D.~Kubiz{\v n}\'ak and R.~B.~Mann, ``Generalized quasitopological gravity,'' Phys. Rev. D \textbf{95}, 104042 (2017) [arXiv:1703.01631 [hep-th]].

\bibitem{Bueno:2017sui}
P.~Bueno and P.~A.~Cano, ``On black holes in higher-derivative gravities,'' Class. Quant. Grav. \textbf{34}, 175008 (2017) [arXiv:1703.04625 [hep-th]].

\bibitem{Bueno:2019ltp}
P.~Bueno, P.~A.~Cano, J.~Moreno and A.~Murcia, ``All higher-curvature gravities as Generalized quasi-topological gravities,'' JHEP \textbf{11}, 062 (2019) [arXiv:1906.00987 [hep-th]].

\bibitem{Quiros:2020uhr}
I.~Quir\'os, R.~Garc\'\i a-Salcedo, T.~Gonzalez, J.~L.~M.~Mart\'\i nez and U.~Nucamendi, ``Global asymptotic dynamics of Cosmological Einsteinian Cubic Gravity,'' arXiv:2003.10516 [gr-qc].

\bibitem{Erices:2019mkd}
C.~Erices, E.~Papantonopoulos and E.~N.~Saridakis, ``Cosmology in cubic and $f(P)$ gravity,'' Phys. Rev. D \textbf{99}, 123527 (2019) [arXiv:1903.11128 [gr-qc]].

\bibitem{Arciniega:2020pcy}
G.~Arciniega, L.~Jaime and G.~Piccinelli, ``Inflationary predictions of Geometric Inflation,'' Phys. Lett. B \textbf{809}, 135731 (2020) [arXiv:2001.11094 [gr-qc]].

\bibitem{Marciu:2020ysf}
M.~Marciu, ``Note on the dynamical features for the extended $f(P)$ cubic gravity,'' Phys. Rev. D \textbf{101}, 103534 (2020) [arXiv:2003.06403 [gr-qc]].

\bibitem{Pookkillath:2020iqq}
M.~C.~Pookkillath, A.~De Felice and A.~A.~Starobinsky, ``Anisotropic instability in a higher order gravity theory,'' JCAP \textbf{07}, 041 (2020) [arXiv:2004.03912 [gr-qc]].

\bibitem{Marciu:2020ski}
M.~Marciu, ``Dynamical aspects for scalar fields coupled to cubic contractions of the Riemann tensor,'' Phys. Rev. D \textbf{102}, 023517 (2020) [arXiv:2004.07120 [gr-qc]].

\bibitem{Camanho:2014apa}
X.~O.~Camanho, J.~D.~Edelstein, J.~Maldacena and A.~Zhiboedov, ``Causality constraints on corrections to the graviton three-point coupling,'' JHEP {\bf 1602}, 020 (2016) [arXiv:1407.5597 [hep-th]].

\bibitem{Camanho:2015bbt}
X.~O.~Camanho, J.~D.~Edelstein and A.~Zhiboedov, ``Weakly coupled gravity beyond general relativity,'' Int.\ J.\ Mod.\ Phys.\ D {\bf 24}, 1544031 (2015).

\bibitem{Lovelock:1971yv}
D.~Lovelock, ``The Einstein tensor and its generalizations,'' J. Math. Phys. \textbf{12}, 498 (1971).

\bibitem{Bueno:2016ypa}
P.~Bueno, P.~A.~Cano, V.~S.~Min and M.~R.~Visser, ``Aspects of general higher-order gravities,'' Phys. Rev. D \textbf{95}, 044010 (2017) [arXiv:1610.08519 [hep-th]].

\bibitem{Hohm:2019jgu}
O.~Hohm and B.~Zwiebach, ``Duality invariant cosmology to all orders in $\alpha$','' Phys. Rev. D \textbf{100}, 126011 (2019) [arXiv:1905.06963 [hep-th]].

\bibitem{Edelstein:2019wzg}
J.~D.~Edelstein, K.~Sfetsos, J.~Sierra-Garcia and A.~Vilar L\'opez, ``T-duality equivalences beyond string theory,'' JHEP \textbf{05}, 082 (2019) [arXiv:1903.05554 [hep-th]].

\bibitem{Hennigar:2018hza}
R.~A.~Hennigar, M.~B.~J.~Poshteh and R.~B.~Mann, ``Shadows, Signals, and Stability in Einsteinian Cubic Gravity,''
Phys. Rev. D \textbf{97}, no.6, 064041 (2018) [arXiv:1801.03223 [gr-qc]].

\bibitem{Khodabakhshi:2020hny}
H.~Khodabakhshi, A.~Giaimo and R.~B.~Mann, ``Einsteinian Quartic Gravity: Shadows, signals and stability,'' Phys. Rev. D \textbf{102}, 044038 (2020) [arXiv:2006.02237 [gr-qc]].

\bibitem{Weinberg:2008zzc}
S.~Weinberg, {\it Cosmology}, Oxford University Press, 2008.

\bibitem{Edelstein:2020lgv}
J.~D.~Edelstein, R.~B.~Mann, D.~V.~Rodr\'\i{}guez and A.~Vilar L\'opez, ``Small free field inflation in higher curvature gravity,'' arXiv:2007.07651 [hep-th].
\end{thebibliography}
\end{document}